\documentclass{fac}

\usepackage{graphicx}
\usepackage{amssymb}
\usepackage{amsfonts}
\usepackage{amsmath}
\usepackage{dsfont}
\usepackage{MnSymbol}
\usepackage{mathtools}
\usepackage{epsfig}
\usepackage{epstopdf}
\usepackage{tikz}
\usepackage{amsthm}
\ifprodtf \else \usepackage{latexsym}\fi

\title[Draft of MA-UNet]
      {MA-Unet: An improved version of Unet based on multi-scale and attention mechanism for medical image segmentation}

\author[Yong Wang]
    {Yutong Cai, Yong Wang\\
     College of Computer Science and Technology,\\
     Faculty of Information Technology,\\
     Beijing University of Technology, Beijing, China\\
     }

\correspond{Yong Wang, Pingleyuan 100, Chaoyang District, Beijing, China.
            e-mail: wangy@bjut.edu.cn}

\pubyear{2019}
\pagerange{\pageref{firstpage}--\pageref{lastpage}}

\begin{document}
\label{firstpage}

\makecorrespond

\maketitle

\begin{abstract}
Although convolutional neural networks (CNNs) are promoting the development of medical image semantic segmentation, the standard model still has some shortcomings. First, the feature mapping from the encoder and decoder sub-networks in the skip connection operation has a large semantic difference. Second, the remote feature dependence is not effectively modeled. Third, the global context information of different scales is ignored. In this paper, we try to eliminate semantic ambiguity in skip connection operations by adding attention gates (AGs), and use attention mechanisms to combine local features with their corresponding global dependencies, explicitly model the dependencies between channels and use multi-scale predictive fusion to utilize global information at different scales. Compared with other state-of-the-art segmentation networks, our model obtains better segmentation performance while introducing fewer parameters.
\end{abstract}


\section{Introduction}

Semantic segmentation of medical images is a key step in the diagnosis, treatment and follow-up of many diseases. In clinical practice, medical image segmentation usually uses manual or semi-manual segmentation technology. The disadvantage of these methods is to use hand-made features to obtain segmentation results. On the one hand, it is difficult to design representative characteristics for different applications. On the other hand, functions designed for one type of image often fail in another type of image. Therefore, traditional manual or semi-manual segmentation techniques lack universal feature extraction methods. Because the manual intensive labeling of a large number of medical images is a tedious and error-prone task, people have put forward higher requirements for accurate and reliable automatic segmentation methods to improve work efficiency in clinical scenarios and reduce the workload of radiologists and other medical experts.

In recent years, convolutional neural networks have achieved the most advanced performance in many image and video vision tasks with their powerful non-linear feature extraction capabilities. The segmentation method of deep learning is based on pixel classification. Unlike traditional pixel or super pixel classification methods, deep learning methods learn features to overcome the limitations of manual methods.

FCN \cite{1} classifies images at the pixel level. Unlike classic convolutional neural networks (CNNs) that use fully connected layers after convolutional layers to obtain fixed-length feature vectors for classification, FCN can accept input images of any size, and use the deconvolution layer to upsample the feature map of the last convolution layer to restore it to the same size as the input image, so that a prediction can be generated for each pixel, while retaining the spatial information in the original input image, thereby solving the problem of image segmentation at the semantic level. Unet \cite{2} draws on the FCN network, and its network structure consists of two parts. The shrinking network of the former part uses 3x3 convolution and pooling downsampling to capture the context information in the image; the extended network of the latter part uses 3x3 convolution and upsampling, in order to achieve the purpose of precise positioning of the required segmentation part of the image. In addition, feature fusion is also used in the network, and down-sampling features of the former part and up-sampling features of the latter part are spliced and fused to obtain more accurate context information and achieve better segmentation results. UNet++ \cite{3} integrates Unet structures of different sizes into a network, captures features of different levels, and integrates them into a shallower Unet structure through feature superposition, so that the scale difference of feature maps during fusion is smaller. Attention U-net \cite{4} proposed a mechanism of attention gates (AGs). Attention gates (AGs) implicitly generate soft region suggestions, highlighting salient features useful for specific tasks. By suppressing the features of irrelevant regions, the sensitivity and accuracy of the model for dense label prediction are improved. In addition, they do not introduce a lot of computational overhead and do not require a lot of model parameters. Channel-UNet \cite{5} is an end-to-end convolutional neural network, which uses Unet as the main structure of the network, and adds spatial channel convolution to each up-sampling and down-sampling module. The network can converge the optimal mapping relationship of spatial information between pixels extracted by spatial channel convolution and information extracted by feature mapping, and realize multi-scale information fusion. In addition, an iterative expansion learning strategy is proposed, which optimizes the mapping relations of spatial information between pixels at different scales, and enables spatial channel convolution to map the spatial information between pixels in high-level feature maps. CE-Net \cite{6} relies on the dense hole convolution module and the residual multi-core pooling module to capture more abstract features and retain more spatial information to improve the performance of medical image segmentation. The dense hole convolution block captures broader and deeper semantic features by injecting four cascaded branches with multi-scale hole convolution, and uses shortcut connections to prevent the problem of gradient disappearance. The residual multi-core pooling block further encodes the multi-scale context features of the target extracted from the dense hole convolution block by adopting pooling operations of various sizes, without additional learning weights. These depth models dominate the field of medical image segmentation. These networks and their extensions have been widely used in these scenarios \cite{7}, \cite{8}, \cite{9}, \cite{10}, \cite{11}.

The main contributions of this paper are summarized as follows:

\begin{enumerate}
  \item In order to improve the performance of medical image segmentation and reduce the complexity of the network structure, this paper proposes a model MA-UNet for semantic segmentation of medical images. This model uses the more lightweight Attention U-net as the basic network architecture. Then the multi-scale and attention mechanism were introduced to obtain the final segmentation results, which achieved better results than the previous UNet series models.
  \item This paper proposes a multi-scale mechanism. This method aggregates features generated by multiple intermediate layers for prediction, and integrates and utilizes global information of different scales.
  \item This paper introduces an attention mechanism to express the dependencies in space and channel dimensions in parallel. This method can establish the association between features and attention mechanism to explore global context information.
  \item Comparing our model with the UNet series models proposed in recent years, the experimental results show that the results of the model in this paper have achieved better results than the previous models, confirming the effectiveness of the method.
\end{enumerate}

\section{Related work}

\subsection{Attention U-Net}

High representation ability, fast reasoning and filter sharing characteristics make Convolutional Neural Networks (CNNs) the best choice for image semantic segmentation. Unet and its variants are commonly used architectures. Although these structures have good expressive power, maintaining high-resolution feature maps in the intermediate stage can improve segmentation performance, but it also increases the size of the feature maps, which is not ideal for accelerating training and reducing optimization difficulties, At the same time, it will lead to excessive and redundant use of computing resources and model parameters; for example, all models in the cascade will repeatedly extract similar low-level features. In order to solve this common problem, Attention U-net \cite{4} demonstrated the realization of attention gates (AGs). Each sub-attention gate (AG) extracts complementary information and merges to define the output of skip connections. The corresponding linear transformation decouples the feature maps and maps them to the low-dimensional space for gating operation, and uses deep supervision \cite{12} to force the intermediate feature maps to be semantically distinguishable at each image scale, which helps to ensure attention units of different scales can respond to the foreground content of a wide range of images. Attention gates (AGs) can automatically learn to focus on target structures of different shapes and sizes. It will implicitly generate soft area suggestions during operation, suppress irrelevant areas in the input image, and highlight salient features useful for specific tasks to eliminate the ambiguity caused by irrelevant and noisy responses in skip connections. In addition, they do not introduce a large amount of computational overhead, and do not require a large number of model parameters like the multi-model framework, and can improve the sensitivity and accuracy of the model for dense label prediction. This model eliminates the necessity of applying external object positioning models, and has the characteristics of versatility and modularity.

\subsection{Attention mechanism}

The basic idea of the attention mechanism in computer vision is to let the system learn to pay attention and be able to ignore irrelevant information and focus on key information. In recent years, attention models have been widely used in various fields such as image processing, speech recognition, and natural language processing. With the development of deep learning today, building a neural network with an attention mechanism is beginning to become more important. On the one hand, this neural network can learn the attention mechanism autonomously, and on the other hand, the attention mechanism can in turn help us understand the world seen by the neural network \cite{13}. In recent years, most of the research work on the combination of deep learning and visual attention mechanism has focused on using masks to form attention mechanisms. The principle of the mask is to identify the key features in the image data through another layer of new weights. Through learning and training, the deep neural network can learn the areas that need attention in each new image. This kind of thinking has evolved into two different types of attention, one is soft attention and the other is hard attention. Hard attention \cite{14} attention point, that is, every point in the image may extend attention. At the same time, hard attention is a random prediction process, which emphasizes dynamic changes. The most important thing is that hard attention is an undifferentiable attention, the training process is often done through reinforcement learning, which makes model training more difficult. The key point of soft attention is that this kind of attention pays more attention to areas \cite{15} or channels \cite{16}, and soft attention is deterministic attention, which can be directly generated through the network after learning is completed. The most important thing is soft attention is differentiable.

In recent years, the development of the soft attention module has made rapid progress. SENet \cite{16} uses the global average pool independently for each channel, and then uses two non-linear fully connected layers and a sigmoid function to generate the weight of each channel. The design of these two fully connected layers is to capture non-linear cross-channel interactions while reducing dimensionality to avoid excessive model complexity. CBAM \cite{17} is a simple and effective attention module for feedforward convolutional neural networks. Given an intermediate feature map, the CBAM module will infer the attention map in turn along the two independent dimensions of channel and space. Then multiply the attention map with the input feature map to perform adaptive feature optimization. CBAM is a lightweight general-purpose module that can be seamlessly integrated into any convolutional neural network architecture without the overhead of the module, and can be trained end-to-end together with the basic convolutional neural network. ECA \cite{18} through the analysis of the channel attention module in SENet, empirically proved that avoiding dimensionality reduction is very important for learning channel attention, and proper cross-channel interaction can significantly reduce model complexity while maintaining performance. Therefore, ECA proposes a local cross-channel interaction strategy without dimensionality reduction, which can be effectively implemented through one-dimensional convolution. In addition, a method for adaptively determining the size of the one-dimensional convolution kernel was developed to determine the coverage of the local cross-channel interaction. This attention module is efficient, involving only a small number of parameters, and it also brings obvious performance gain. GSoP-Net \cite{19} performs covariance matrix calculation, and then performs two consecutive operations of linear convolution and nonlinear activation to obtain the output tensor. The output tensor scales the original input along the channel dimension, which is also a manifestation of channel attention to a certain extent. But unlike SEnet, this module proposes two-dimensional average pooling, which embodies the relationship between channels in the form of covariance, and more effectively aggregates features. Inspired by the classic non-local means \cite{20} in computer vision, the proposed Non-local \cite{21} is used to capture long-distance dependence, that is, how to establish the relationship between two pixels with a certain distance on the image. When non-local calculates the response of a certain location, it considers the weighting of all location features. All locations can be spatial, temporal, or even spatial-temporal. In addition, this attention module can be inserted into many computer vision structures. Residual Attention Network \cite{22} is a convolutional neural network that uses an attention mechanism, which can be combined with the latest feedforward network structure in an end-to-end training method. The Residual Attention Network is formed by stacking attention modules, which generate attention perception features. As the level deepens, the attention perception characteristics from different modules will adaptively change. In each attention module, the bottom-up and top-down feedforward structure expands the feedforward and feedback attention process into a single feedforward process. It is also proposed that attention residual learning is very important for training a very deep Residual Attention Network, which can be easily extended to hundreds of layers. DANet \cite{23} proposed a dual attention network to adaptively combine local features with their global correlation. On the basis of the expanded FCN, two types of attention modules are added to capture the feature dependence in the spatial dimension and the channel dimension respectively. The spatial attention module uses a self-attention mechanism to capture the spatial dependence of the feature map between any two positions, and aggregates and updates the features of all positions through weighted summation. The weight is determined by the similarity of the features corresponding to the two positions. At the same time, the channel attention module uses a self-attention mechanism to capture the channel dependency between any two channel graphs, and updates each channel graph with the weight of all the channel graphs. The outputs of the two attention modules are added together to further improve the feature representation, thereby selectively aggregating similar features of inconspicuous objects, highlighting their feature representations, avoiding the influence of prominent objects, and adaptively integrating similar features on any scale from a global perspective. CCNet \cite{24} can effectively obtain the context information of the complete image. For each pixel, a novel cross-attention module obtains the context information of all pixels on its cross path. Through further loop operations, each pixel can finally capture the complete image dependency. In addition, a category consistency loss is proposed to strengthen the cross-attention module to produce more distinguishing features. CCNet not only occupies less GPU memory, but also has high computational efficiency. The segmentation performance is also significantly improved. GCNet \cite{25} not only can effectively model remote dependencies, but also has the characteristics of lightweight. In the main benchmark tests of various recognition tasks, its performance is usually better than the simplified Non-local and SEnet.

\subsection{Multi-scale output}

Multi-scale is actually sampling the different granularities of the signal. Usually, different characteristics can be observed at different scales to complete different tasks. In computer vision, scale is always a big issue, and small objects and very large-scale objects often seriously affect performance. The convolutional neural network extracts the features of the target through a layer-by-layer abstraction. One of the important concepts is the receptive field. If the receptive field is too small, only local features can be observed. If the receptive field is too large, too much invalid information will be obtained. Therefore, researchers have been designing various multi-scale model architectures. The network structure can be divided into multi-scale input, multi-scale feature fusion, multi-scale feature prediction fusion, and a combination of these three structures. Multi-scale input is to use multiple scales of image input, and then fuse the results. The traditional MTCNN \cite{26} face detection algorithm uses this idea. There are two common multi-scale feature fusion networks. The first is a parallel multi-branch network, and the second is a serial layer-jumping connection structure. Both of them extract features in different receptive fields. Multi-scale feature prediction fusion is used to predict at different feature scales, and finally the results are fused, represented by SSD \cite{27} in target detection.

\section{Method}

\subsection{Overview of MA-Unet}

Most of the main network architectures involved in medical image semantic segmentation tasks have a key idea: skip connection. The most commonly used network architecture, Unet, connects them in series and adds convolution and nonlinear operations between each upsampling step. The skip connection helps restore the full spatial resolution at the output of the network, making the fully convolutional method suitable for semantic segmentation. However, the feature mapping from the encoder and decoder sub-networks has a large semantic difference, and this semantic difference will increase the learning difficulty of the network and reduce the segmentation performance. Secondly, in two-dimensional image processing, traditional convolutional neural network segmentation has local receptive fields, thereby generating local feature representations. Although learning context dependence on local features can also help feature representation, the lack of long-term dependence and the inability to make full use of the object-to-object relationship in the global view may lead to potential differences between corresponding features of pixels with the same label. At the same time, these networks do not make full use of the feature information of the intermediate convolutional layer, and ignore the global context information of different scales.

\begin{figure}
    \centering
    \includegraphics[width=16cm,height=6cm]{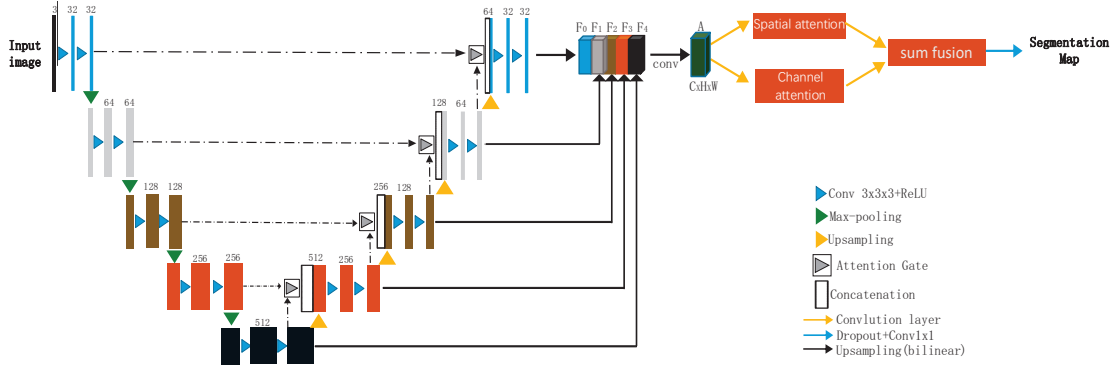}
    \caption{Overview of MA-Unet network.}
    \label{f1}
\end{figure}

In response to this problem, we studied the attention mechanism that establishes associations between features and introduced multi-scale prediction fusion. First, the attention gate mechanism (AGs) is incorporated into the Unet architecture to eliminate the ambiguity caused by the irrelevance and noise response in the skip connection, and only merge the relevant activations. Then, establish the association between the feature and the attention mechanism to explore the global context information, establish the channel attention module to explicitly model the dependency between the channels, and then establish the spatial attention module to encode the broader context information into local features, improving the representation ability of local features, and aggregate the features of these two attention modules. Among them, the spatial attention module combines the advantages of Non-local and SEnet, which can not only establish effective long-distance dependence like Non-local, but also save computation like SEnet, and has the advantage of being lighter. Finally, the multi-scale prediction is merged into the architecture, and the features generated by multiple intermediate layers are aggregated for prediction, so as to utilize global information of different scales. Fig. \ref{f1} depicts an overview of the proposed framework.

We use the more lightweight Attention U-net as the basic network architecture. Then we aggregate the features generated by multiple intermediate layers for prediction, using global information at different scales. Finally, the association between the feature and the attention mechanism is established to explore the global context information. The attention module will generate attention features on multiple scales, remove noise areas, and help the network emphasize areas more related to semantic classes.

\subsection{Spatial and channel attention mechanism model}

\begin{figure}
    \centering
    \includegraphics{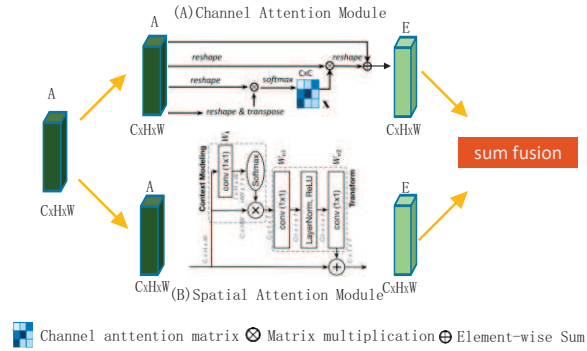}
    \caption{Details of spatial attention module and channel attention module are shown in (A) and (B).}
    \label{f2}
\end{figure}

As mentioned earlier, in the traditional segmentation depth model, the receptive field is reduced to the local vicinity. This limits the ability to represent a broader and richer context. As shown in Fig. \ref{f2}, we introduce an attention mechanism to express the dependencies in the space and channel dimensions in parallel.

\subsubsection{Channel attention mechanism model}

Each channel graph of high-level features can be regarded as a class specific response, and different semantic responses are related to each other. Using the interdependence between channel graphs, we can highlight the feature mapping of interdependence and improve the feature representation of specific semantics. Therefore, we establish a channel attention module to explicitly simulate the interdependence between channels and obtain the correlation strength of any two channels.

The structure of the channel attention module is shown in Fig. \ref{f2}(A). We directly calculate the channel attention map $X\in R^{C\times C}$ from the original features $A\in R^{C\times H\times W}$. Specifically, we reshape $A$ to $R^{C\times N}$, and then perform a matrix multiplication between $A$ and the transpose of $A$. Finally, we apply a softmax layer to obtain the channel attention map $X\in R^{C\times C}$:

\begin{equation}\label{e1}
x_{ij}=\frac{exp(A_i.A_j)}{\sum^C_{i=1}exp(A_i.A_j)}
\end{equation}

Where $x_{ij}$ represents the influence of the $i^{th}$ channel on the $j^{th}$ channel. In addition, we perform a matrix multiplication between the transpose of $X$ and $A$ and reshape their result to $R^{C\times H\times W}$. Then we multiply the result by a scale parameter $\beta$ and perform an element-wise sum operation with $A$ to obtain the final output $E\in R^{C\times H\times W}$:

\begin{equation}\label{e2}
E_j=\beta\sum^C_{i=1}(x_{ji}A_i)+A_j
\end{equation}

$\beta$ gradually learns the weight from 0. Equation \ref{e2} shows that the final feature of each channel is the weighted sum of the features of all channels and the original features. It simulates the long-term semantic dependency between feature maps, helps to improve the distinguishability of features, highlights the class-related feature mapping and increases the feature distinguishability between classes.

\subsubsection{Spatial attention mechanism model}

In order to build a rich context model based on local features, a spatial attention module is introduced. The spatial attention module encodes broader contextual information into local features, thereby enhancing their expressive ability. Since the global context of Non-local is almost the same in different locations, it shows that Non-local has learned the global context without position dependence. Using this discovery to simplify Non-local by calculating a global map, and sharing this global map for all locations, which not only maintains almost the same accuracy as Non-local, but also significantly reduces the amount of calculation.

The structure of the space attention module is shown in Fig. \ref{f2}(B). It has the advantages of simplified Non-local and lightweight squeeze and excitation block. Since the two-layer bottleneck transform increases the difficulty of optimization, we added layer normalization in the bottleneck transform before the ReLU operation. The layer normalization can significantly enhance object detection and instance segmentation to simplify optimization, and as a regularizer, it is conducive to generalization . The module is expressed in detail as:

\begin{equation}\label{e3}
Z_i=A_i+W_{v2}ReLU(LN(W_{v1}\sum^{N_p}_{j=1}\frac{e^{W_kA_j}}{\sum^{N_p}_{m=1}e^{W_kA_m}}A_j))
\end{equation}

We denote $A=\{A_i\}_(i=1)^(N_p )$ as a feature map of an input instance, where $N_p$ is the number of positions in the feature map,  and $A$ and $Z$ respectively represent the input and output. Where $\frac{e^{W_k A_j}}{\sum_me^{W_k A_m}}$ is the weight of global attention concentration, and $\delta(.)=W_{v2}ReLU(LN(W_{v1}(.)))$ represents bottleneck transformation. At the end of the two attention modules, the newly generated features perform element-wise summation to generate new features.

\subsection{Multi-scale output}

\begin{figure}
    \centering
    \includegraphics{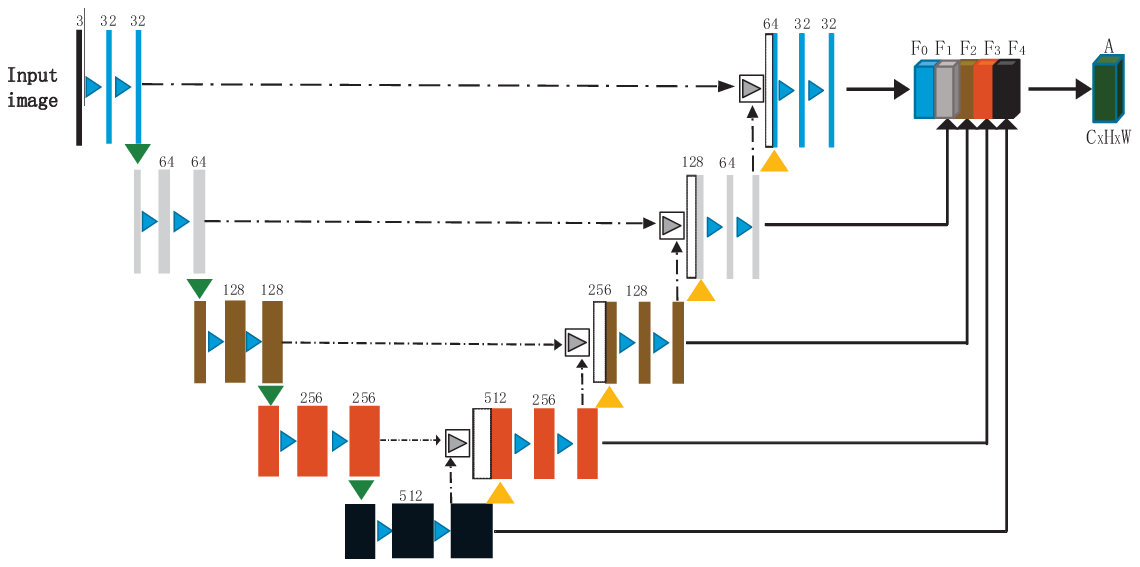}
    \caption{The structure of a multi-scale module.}
    \label{f3}
\end{figure}

Before the rapid development of deep learning, multi-scale features have been widely used in computer vision problems \cite{28}, and the fusion of multi-scale features has also shown amazing performance \cite{29}, \cite{30}. Using learned features at multiple scales helps to encode global and local contexts. The structure of the multi-scale module is shown in Fig. \ref{f3}. Specifically, the multi-scale feature prediction fusion is denoted as A. Since the features of each scale have different resolutions, they are up-sampled to a common resolution through bilinear interpolation. Then, feature maps from all scales are concatenated to form a tensor, which is convolved to create a multi-scale feature prediction fusion $A=conv([F_0,F_1,F_2,F_3])$. Therefore, $A$ coding comes from the low-level details of the shallow layer and the high-level semantics learned in the deep layer.

\subsection{Loss function}

This article uses Binary Cross-Entropy as the objective function. Binary cross entropy is the cross entropy of two categories, which is actually a special case of multi-category cross entropy. Suppose $Pre_i$ represents the image predicted by the $i^{th}$ model, $GT_{i}$ represents the Ground Truth of the $i^{th}$ image, and $GT_{i,j}$ represents the pixel value at the $i^{th}$ and $j^{th}$ positions of the Ground Truth, $Pre_{i,j}$ represents the pixel value of the  $i^{th}$ and  $j^{th}$ positions of the model predicted image, $W$ represents the width of the image, and $H$ represents the height of the image. Then the binary cross entropy loss can be expressed as:

\begin{equation}\label{e4}
L_{BCE}=BCELoss(Pre_i,GT_i)=-\sum^W_{i=1}\sum^H_{j=1}[(GT_{i,j}\times logPre_{i,j})+(1-GT_{i,j})\times log(1-Pre_{i,j})]
\end{equation}

\section{Experiment}

In order to evaluate the method proposed in the article, we conducted comprehensive experiments on the lung segmentation dataset and the esophagus and esophagus cancer dataset from First Affiliated Hospital of Sun Yat-sen University. The experimental results show that the method proposed in the article achieves the most advanced performance on both data sets and introduces fewer parameters of the network structure. In the following chapters, we first introduce the lung segmentation dataset and the esophagus and esophagus cancer dataset from First Affiliated Hospital of Sun Yat-sen University, implementation details and evaluation indicators, and then conduct a series of experiments on these two datasets. Finally, we report the experimental results.

\subsection{Datasets}

\subsubsection{Lung segmentation dataset}

It is applied to the task of lung segmentation, which is to segment the lung structure in the 2D CT image from the lung nodule analysis (LUNA) competition. The LUNA competition was originally conducted for the following challenge tracks: nodule detection and false positive reduction. Because segmented lungs are the basis for further lung nodule candidates, we use challenge data sets to evaluate our proposed network structure MA-Unet. The data set contains 534 two-dimensional samples ($512\times 512$ pixels) and their respective label images. We use 80\% of the images for training and the rest for testing.

\subsubsection{The esophagus and esophagus cancer dataset from First Affiliated Hospital of Sun Yat-sen University}

It is applied to the task of esophageal and esophageal cancer segmentation, the data set includes 13,239 two-dimensional samples ($80\times 80$ pixels) and their respective label images for training, including 2949 two-dimensional samples ($80\times 80$ pixels) and their respective label images for testing.

\subsection{Experimental details and evaluation indicators}

\subsubsection{Experimental details}

We use the Python3 programming language for experiments, and the network model is implemented using the Pytorch backend. We use the Adam optimizer with a learning rate of 0.001 to train all networks. The batch size of the lung segmentation dataset is set to 1, the number of training iterations is set to 50, and the batch size of the esophagus and esophagus cancer dataset from First Affiliated Hospital of Sun Yat-sen University is set to 8, the number of training iterations is set to 100. The loss functions we use in all networks of all categories are binary cross-entropy errors. Evaluation indicators include Mean Intersection over Union and Mean Dice coefficient of common indicators for medical image segmentation. In addition, we added the size of the parameters introduced by the network structure to compare the complexity of the network structure. And after each round of training, the model in this round will be compared with the current model, if it is better than the current model, it will be saved, otherwise it will be ignored.

\subsubsection{Evaluating indicator}

There are two performance evaluation indicators used in this experiment, one is the Mean Intersection over Union (MIOU), which is most commonly used in segmentation tasks, and the other is the Mean Dice coefficient(MDC). They are used to assist in evaluating the quality of different models.
the Mean Intersection over Union is the most common indicator for the performance evaluation of semantic segmentation models. It refers to the overlap rate of the generated candidate frame and the original labeled frame, that is, the ratio of their intersection and union. Assume that $p_{ii}$  is the number of predicted correct elements, $p_{ij}$ represents the true value of $i$, the number of predicted $j$, $p_{ji}$ represents the true value of $j$, the predicted number of $i$, and $k$ represents the number of categories to be classified. Then the Mean Intersection over Union can be expressed as:

\begin{equation}\label{e5}
MIOU=\frac{1}{k+1}\sum^k_{i=0}\frac{p_{ij}}{\sum^k_{j=0}p_{ij}+\sum^k_{j=0}p_{ji}-p_{ii}}
\end{equation}

The Mean Dice coefficient is a measure function of set similarity, which can be used to calculate the similarity between the segmentation map and Ground Truth. Suppose $Pre_i$ represents the segmentation result of the $i^{th}$ image, $GT_i$ represents the Ground Truth of the $i^{th}$ image, and $N$ represents the number of samples. Then the Mean Dice coefficient can be expressed as:

\begin{equation}\label{e6}
DiceLoss(Pre_i,GT_i)=\frac{\sum^N_{i=1}|Pre_i\cap GT_i|}{\sum^N_{i=1}|Pre_i|\cap|GT_i|}
\end{equation}

\subsection{Experimental results}

This section compares our proposed model with existing models to verify the effectiveness of this method. We choose the best models in the medical image segmentation task for comparison, namely Attention U-net, CE-Net, UNet++, Unet and Channel-UNet. Table \ref{t1} and Table \ref{t2} show the corresponding results, and Fig. \ref{f4} and Fig. \ref{f5} show the corresponding segmentation examples.

\subsubsection{Results on Lung segmentation dataset}

\begin{center}
    \begin{table}
    \begin{tabular}{|c|c|c|c|}
        \hline
        Method & MIOU(\%) & MDC(\%) & Parameter(MB)\\
        \hline
        Attention U-net & 95.34 & 97.29 &	133\\
        \hline
        CE-Net & 94.95 & 97.02 & 110\\
        \hline
        UNet++ & 95.21 & 97.22 & 35\\
        \hline
        Unet & 95.03 &	97.12 &	29.6\\
        \hline
        Proposed & 95.76 &	97.52 &	34.5\\
        \hline
    \end{tabular}
    \caption{Comparison of lung segmentation detection performance and parameters.}
    \label{t1}
    \end{table}
\end{center}

Table \ref{t1} reports the experimental results obtained by several state-of-the-art segmentation networks on the lung segmentation dataset. From the results in Table \ref{t1}, it can be seen that the MIOU obtained by Attention U-net, CE-Net, UNet++, and Unet are 95.34\%, 94.95\%, 95.21\%, and 95.03\%, respectively; the MDC obtained are 97.29\%, 97.02\%, 97.22\%, and 97.12\%, respectively. The MIOU and MDC of MA-Unet were 95.76\% and 97.52\%, respectively, and the highest MIOU and MDC were obtained. Compared with the results of the other 4 models, MA-Unet improved by 0.42\%, 0.81\%, 0.55\% and 0.73\% (MIOU), 0.23\%, 0.5\%, 0.3\% and 0.4\% (MDC), respectively. This shows that MA-Unet has achieved better results than other models, reaching state-of-the-art. It can also be seen that our proposed method only increases the parameter amount of 4.9MB compared to the basic model Unet, while the performance is better than the second-performing Attention U-net, and even lower than its parameter amount by 98.5MB.

\subsubsection{Results on the esophagus and esophagus cancer dataset from First Affiliated Hospital of Sun Yat-sen University}

\begin{center}
    \begin{table}
    \begin{tabular}{|c|c|c|c|}
        \hline
        Method & MIOU(\%) & MDC(\%) & Parameter(MB)\\
        \hline
        Attention U-net & 64.32 & 73.93 &	133\\
        \hline
        Channel-UNet & 64.55 & 74.53 & 187\\
        \hline
        UNet++ & 52.32 & 62.93 & 35\\
        \hline
        Unet & 62.73 &	73.56 &	29.6\\
        \hline
        Proposed & 65.3 &	75.49 &	34.5\\
        \hline
    \end{tabular}
    \caption{Comparison of segmentation detection performance and parameter values of esophagus and esophageal cancer.}
    \label{t2}
    \end{table}
\end{center}

Table \ref{t2} reports the experimental results obtained by several state-of-the-art segmentation networks on the esophagus and esophagus cancer dataset of the First Affiliated Hospital of Sun Yat-sen University. From the results in Table \ref{t2}, it can be seen that the MIOU obtained by Attention U-net, Channel-UNet, UNet++, and Unet are 64.32\%, 64.55\%, 52.32\%, and 62.73\%, respectively; the MDC obtained are 73.93\%, 74.53\%, 62.93\%, and 73.56\%, respectively. The MIOU and MDC of MA-Unet were 65.3\% and 75.49\%, respectively, and the highest MIOU and MDC were obtained. Compared with the results of the other 4 models, MA-Unet improved by 0.98\%, 0.75\%, 12.98\% and 2.57\% (MIOU), 1.56\%, 0.96\%, 12.56\% and 1.93\% (MDC), respectively. This shows that MA-Unet has achieved better results than other models, reaching state-of-the-art. It can also be seen that our proposed method only increases the parameter amount of 4.9MB compared to the basic model Unet, while the performance is better than the second-performing Attention U-net, and even lower than its parameter amount by 98.5MB.

This difference in performance can be explained by the fact that the attention module in general computer vision tasks attracts more attention, resulting in more refined strategies, and usually better segmentation results. At the same time, the adopted multi-scale architecture has been sampled at different granularities. The low-level features have higher resolution and contain more position and detailed information. However, due to less convolution operations, it has lower semantics and more noise. High-level features have stronger semantic information, but the resolution is very low, and the perception of details is poor. Combining the two efficiently, taking the advantages and discarding the bad ones, will significantly improve the segmentation task. These differences show that the dual attention mechanism of channel and space and multi-scale mechanism can actually improve the performance of segmentation networks.

\subsection{Visual inspection feature map}

\begin{figure}
    \centering
    \includegraphics{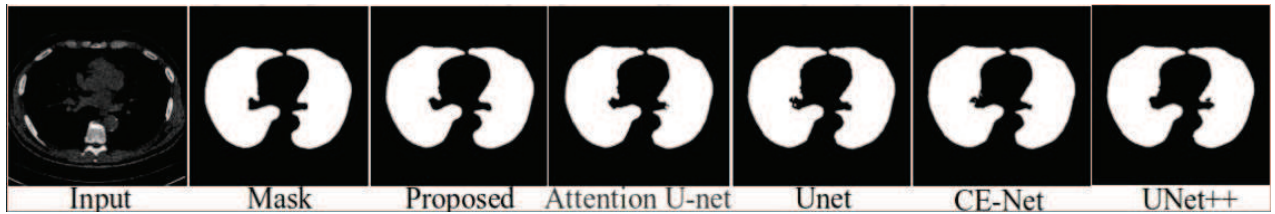}
    \caption{Results of different networks on the lung segmentation dataset.}
    \label{f4}
\end{figure}

\begin{figure}
    \centering
    \includegraphics[width=18cm]{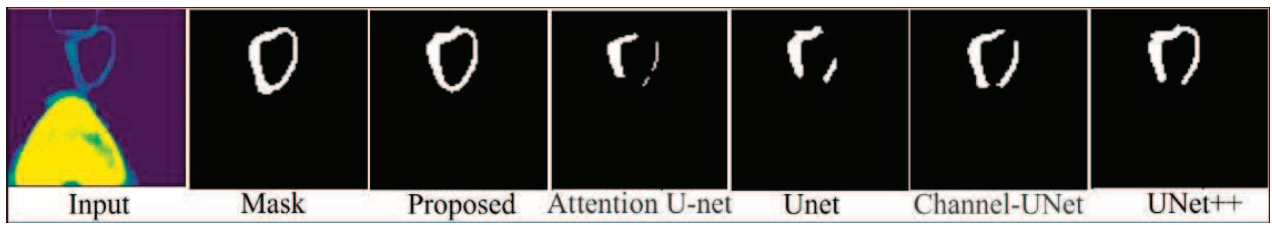}
    \caption{Results of different networks on esophagus and esophagus cancer segmentation datasets.}
    \label{f5}
\end{figure}

Only quantitative evaluation to show performance differences may not be enough to fully understand the advantages and behavior of the proposed model. Although the proposed module contributes to performance improvement, as shown in the results of Table \ref{t1} and Table \ref{t2}, it is also necessary to determine whether the proposed model works as expected through visual observation. For this reason, in Fig. \ref{f4} and Fig. \ref{f5}, we also give some visual comparison examples of lung segmentation and esophagus and esophagus cancer segmentation.

The proposed network combining the attention mechanism and the multi-scale mechanism achieves better results than other latest segmentation networks.

These visual results show that our proposed method can successfully recover finer segmentation details. The multi-scale volume mechanism performs sampling at different granularities, integrates spatial attention and channel attention, and captures contextual information, which can effectively encode complementary information, thereby accurately segmenting medical images.

\section{Conclusions}

Medical image segmentation is an important part of medical image analysis. In this work, we start with a careful analysis of the U-Net architecture, hoping to find potential room for improvement. We found some differences between the features passed from the encoder network and the features passed through the decoder network. In order to coordinate these two sets of incompatible features, we propose to add attention gates (AGs) between the two to extract complementary information and merge them to make the two feature maps more uniform. In addition, in order to use global information at different scales, we propose a multi-scale prediction fusion mechanism. Finally, we established the association between features and spatial attention and channel attention mechanisms, effectively capturing remote context information. Our proposed network architecture MA-Unet has achieved excellent performance on the lung segmentation dataset and the esophagus and esophagus cancer dataset from First Affiliated Hospital of Sun Yat-sen University and introduced a low amount of parameters. It is very important to reduce the computational complexity of the model and enhance the accuracy and robustness of the model. The results obtained by the network architecture mentioned in the article are also closer to the actual situation visually, and can detect the most subtle boundaries.

We hope that in future experiments, the optimal hyperparameter set of the model can be determined in more detail. In addition, we will evaluate the performance of our model on medical images from other modes, and further improve the accuracy and robustness of our model, and bring better segmentation methods for diversified application development.

\newpage

%

\label{lastpage}

\end{document}